\documentclass{tufte-handout}

\usepackage{amsmath}
\usepackage{graphicx}
\usepackage[numbers]{natbib}
\usepackage[mathscr]{eucal}
\usepackage{algpseudocode}
\usepackage{graphicx}
\usepackage{caption}
\usepackage{subcaption}

\setkeys{Gin}{width=\linewidth,totalheight=\textheight,keepaspectratio}
\graphicspath{{graphics/}}

\title[Tesselating Pascal's tetrahedron]{Tesselating a Pascal-like tetrahedron for the\\ {subdivision} of high order tetrahedral finite elements}

\author{Mark W. Lohry}

\date{October 2021}  

\usepackage{booktabs}

\usepackage{units}

\usepackage{fancyvrb}
\fvset{fontsize=\normalsize}

\usepackage{multicol}

\usepackage{lipsum}

\begin{document}

\maketitle

\begin{abstract}
\noindent
\smallcaps{abstract}\\
\noindent
Three-dimensional $N^{th}$ order nodal Lagrangian tetrahedral finite elements ($P_N$ elements) can be generated using Pascal's tetrahedron $\mathcal{H}$ where each node in 3D element space corresponds to an entry in $\mathcal{H}$. For the purposes of visualization and post-processing, it is desirable to ``subdivide'' these high-order tetrahedral elements into sub-tetrahedra which cover the whole space without intersections and without introducing new exterior edges or vertices. That is, the exterior triangulation of the element should be congruent with the ``natural'' triangulation of the $2D$ Pascal's triangle. This work attempts to describe that process of subdivision for arbitrary $N$.
\end{abstract}

\section{Subdivision}

Claimed without proof:
\begin{itemize}
    \item A $P_N$ tetrahedral element can be subdivided into $N^3$ sub-tetrahedra of equal volume and without adding any exterior edges or vertices. See Polster\cite{polster2004qed,mathologertriangularsquares}  for the 2D ``square triangles'' decomposition into $N^2$ sub-triangles.
    \item Considering the build up of the connectivity by adding another level of Pascal's tetrahedron (corresponding to increasing the order of the element), each additional level $N$ adds $N^3-(N-1)^3= 3N^2+3N+1$ subdivisions.
    \item Assuming level $1$ to have a geometric representation as a right-angled trirectangular tetrahedron with side length $1$ giving a volume $V=1/6$, all of the sub-tetrahedra generated by the following method also have volume $1/6$.
\end{itemize}

\begin{figure}[h!]
\centering
\includegraphics[width=0.6\textwidth,trim={0.5cm 0 1cm 0},clip]{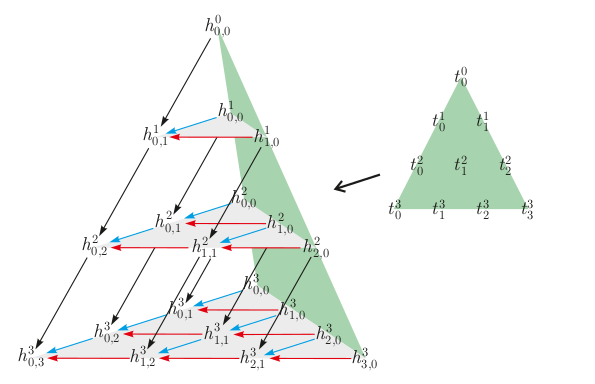}
\caption{Pascal's tetrahedron from N{\'e}meth with indexing convention $h^i_{j,k}$.}
\label{fig:pascaltet}
\end{figure}

\clearpage

\section{Connectivity for the $P_1$ and $P_2$ cases}

The following uses the indexing notation from \cite{nemeth} for the $P_N$ element depicted in figure~\ref{fig:pascaltet}
\begin{equation*}
h^i_{j,k}\ \textrm{where}\quad 0\leq k \leq i,\quad 0\leq j \leq i-k,\quad 0 \leq i \leq N
\end{equation*}

\subsection{The $P_1$ element}

  \begin{marginfigure}
   \includegraphics[width=\linewidth,trim={5cm 0 25cm 0},clip]{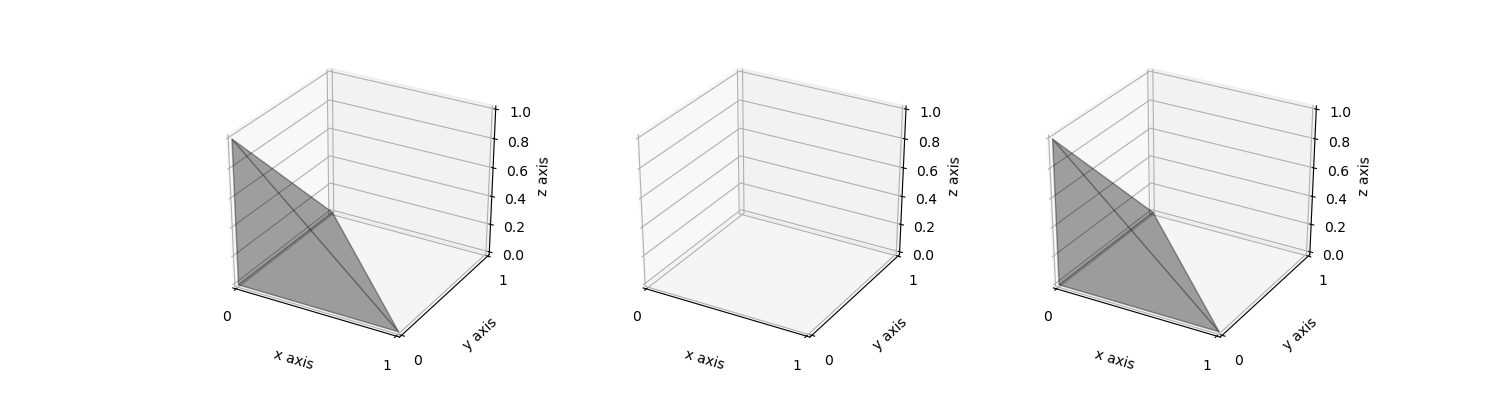}
   \caption{$P_1$ subdivision.}
   \label{fig:marginfig}
 \end{marginfigure}


The only possible $P_1$ ``subdivision'' is 
\begin{equation}
\begin{split}
P_1 &: \{\quad  h^0_{0,0},\quad h^1_{0,0},\quad h^1_{1,0}, \quad h^1_{0,1}  \quad \}
\end{split}
\end{equation}
\setlength\marginparpush{20pt}
\subsection{The $P_2$ element}

For the $P_2$ case\cite{zhang1995successive,vanderzee2008triangulation} translating the $P_1$ row to the three lower corners gives three upright tetrahedra
\begin{align}
P_2 &: \{\quad  h^1_{0,0},\quad h^2_{0,0},\quad h^2_{1,0}, \quad h^2_{0,1}  \quad \} \\
    &: \{\quad  h^1_{1,0},\quad h^2_{1,0},\quad h^2_{2,0}, \quad h^2_{1,1}  \quad \} \\
    &: \{\quad  h^1_{0,1},\quad h^2_{0,1},\quad h^2_{1,1}, \quad h^2_{0,2}  \quad \} 
    \label{eq:uprighttets}
\end{align}
\noindent
Maintaining the exterior face symmetries \sidenote{the exterior faces should look like the simple triangulation of a 2D triangular element} requires additional elements

\begin{align*}
 &: \{\quad  h^1_{0,0},\quad h^1_{1,0},\quad h^2_{1,0}, \quad h^2_{1,1}\ or\ h^2_{0,1}\ or\ h^1_{0,1}  \quad \} \\
 &: \{\quad  h^1_{1,0},\quad h^1_{0,1},\quad h^2_{1,1}, \quad h^2_{1,0}\ or\ h^2_{0,1}\ or\ h^1_{0,0}  \quad \} \\
 &: \{\quad  h^1_{0,0},\quad h^1_{0,1},\quad h^2_{0,1}, \quad h^2_{1,1}\ or\ h^2_{1,0}\ or\ h^1_{1,0}  \quad \} \\
 &: \{\quad  h^2_{1,0},\quad h^2_{0,1},\quad h^2_{1,1}, \quad h^1_{0,0}\ or\ h^1_{1,0}\ or\ h^1_{0,1}  \quad \} 
\end{align*}
\noindent
where the exterior faces formed by the first three nodes are fixed by the symmetry constraint and the last node is a ``free'' choice of the other non-coplanar nodes in the interior. This gives the $2^3=8$ sub-tetrahedra. 

\begin{figure*}
         \centering
         \includegraphics[width=\textwidth,trim={5cm 0 3.5cm 0},clip]{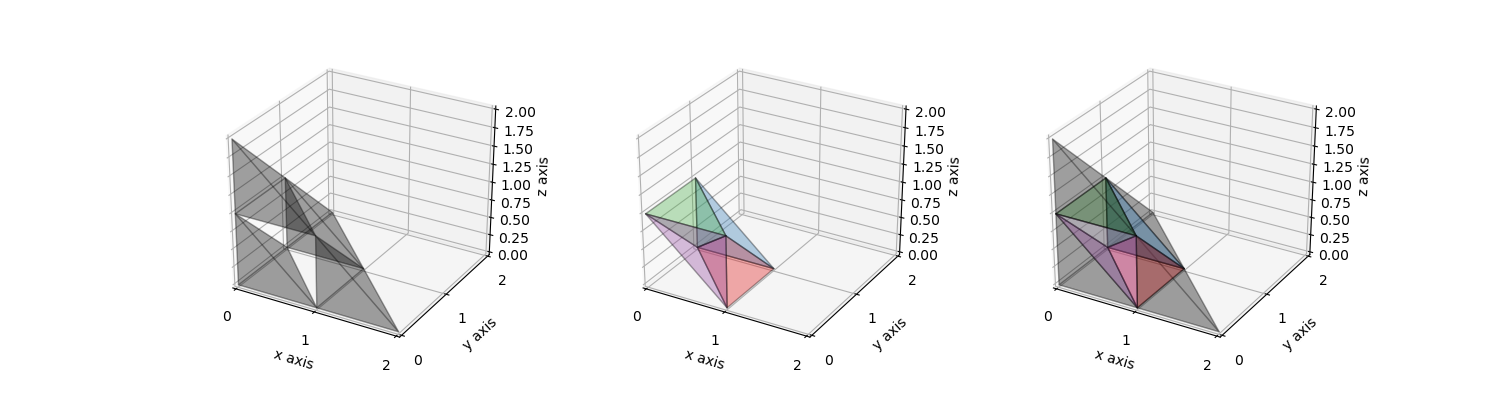}
         \caption{$P_2$ subdivision. }
         \label{fig:p2sub}
\end{figure*}

The interior faces formed by the 3 new interior vertical sub-tetrahedra are
\begin{align*}
    \{\quad h^1_{0,1},\quad h^2_{0,1},\quad h^2_{1,1}\quad \} \\
    \{\quad h^1_{0,0},\quad h^2_{1,0},\quad h^2_{0,1}\quad \} \\
    \{\quad h^1_{0,0},\quad h^2_{1,0},\quad h^2_{1,1}\quad \} \\
\end{align*}
and the horizontal upper and lower triangular faces are
\begin{align*}
    \{\quad h^1_{0,0},\quad h^1_{1,0},\quad h^1_{0,1}\quad \} \\
    \{\quad h^2_{1,0},\quad h^2_{0,1},\quad h^2_{1,1}\quad \} \\
\end{align*}
which forms a regular octahedron\sidenote{8 triangular faces, 6 vertices, 12 edges} as the interior ``hole'' to be subdivided. A new interior edge must be introduced to split this octahedron. Choosing that new edge to be\sidenote{One could instead choose a new edge $\{h^1_{1,0}, h^2_{0,1} \}$ and arrive at a similar but differently oriented subdivision.}
\begin{align*}
    \{\quad h^1_{0,0},\quad h^2_{1,1}\quad \} \\
\end{align*}
leads to the interior decomposition:
\begin{align}
 &: \{\quad  h^1_{0,0},\quad h^2_{1,1},\quad h^1_{1,0}, \quad  h^2_{1,0} \quad \} \\
 &: \{\quad  h^1_{0,0},\quad h^2_{1,1},\quad h^1_{1,0}, \quad  h^1_{0,1} \quad \} \\
 &: \{\quad  h^1_{0,0},\quad h^2_{1,1},\quad h^2_{0,1}, \quad  h^1_{0,1} \quad \} \\
 &: \{\quad  h^1_{0,0},\quad h^2_{1,1},\quad h^2_{0,1}, \quad  h^2_{1,0}  \quad \} 
\end{align}
This is depicted in figure~\ref{fig:p2sub} where this middle subdivision fills the gap left behind by the upright oriented right tetrahedra.

\clearpage

\section{The $P_3$ element and the suggestion of extension to arbitrary order}

The construction of the upright tetrahedra follow from the $P_2$ cells~\ref{eq:uprighttets}. At level $i=N$, $N(N+1)/2$ upright tetrahedra are added

\begin{algorithmic}
\For{$k \gets 0,i-1$ }
\For{$j \gets 0,i-k-1$}
\State New subtet $\gets \{\quad h^{i-1}_{j,k},\quad h^{i}_{j,k},\quad h^{i}_{j+1,k},\quad h^{i}_{j,k+1} \}$
\EndFor
\EndFor
\end{algorithmic}

\noindent
To fill the holes, $2N(N-1)$ tetrahedra are added to each

\begin{algorithmic}
\For{$k \gets 0,i-2$ }
\For{$j \gets 1,i-k-1$}

\State New subtet $\gets \{\quad  h^{i-1}_{j-1,k},\quad h^i_{j,k+1},\quad h^{i-1}_{j,k}, \quad  h^i_{j,k} \quad \}$
\State New subtet $\gets \{\quad  h^{i-1}_{j-1,k},\quad h^i_{j,k+1},\quad h^{i-1}_{j,k}, \quad  h^{i-1}_{j-1,k+1} \quad \}$
\State New subtet $\gets \{\quad  h^{i-1}_{j-1,k},\quad h^i_{j,k+1},\quad h^{i}_{j-1,k+1}, \quad  h^{i-1}_{j-1,k+1} \quad \}$
\State New subtet $\gets \{\quad  h^{i-1}_{j-1,k},\quad h^i_{j,k+1},\quad h^{i}_{j-1,k+1}, \quad  h^{i}_{j,k} \quad \}$
 
\EndFor
\EndFor
\end{algorithmic}


 \begin{figure*}
         \centering
         \includegraphics[width=\textwidth,trim={5cm 0 3.5cm 0},clip]{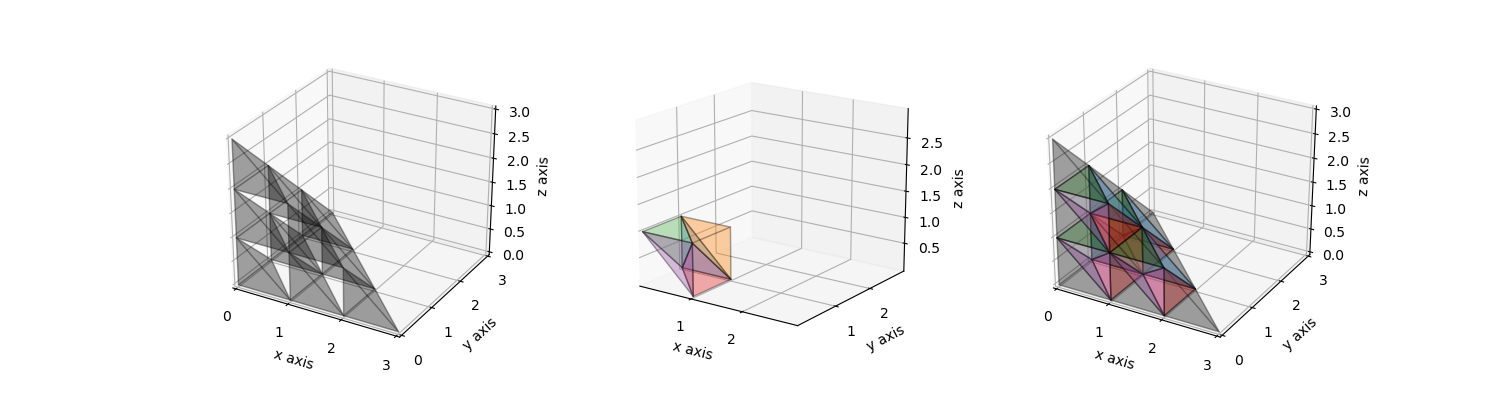}
         \caption{$P_3$ subdivision with innermost gap shown in the center image. }
         \label{fig:p3sub}
\end{figure*}

\clearpage

\section{The $P_4$ element and extension to arbitrary order}

The above misses one sub-tetrahedron construction present at orders $P_3$ and above, where the face $\{h^{i-1}_{j+1,k}, h^{i-1}_{j,k+1}, h^i_{j+1,k+1}\}$ is on the exterior slanted face for $N<3$ but is interior at $N>=3$. Examining the construction of the $3^\text{rd}$ base level with the above algorithm leads to a missing chunk for the interior cell subdivision where for $i \ge 3$ corresponding to
\begin{algorithmic}
\For{$k \gets 0,i-3$ }
\For{$j \gets 0,i-k-3$}
\State New subtet $\gets \{\quad  h^{i-1}_{j+1,k},\quad h^{i-1}_{j+1,k+1},\quad h^{i-1}_{j,k+1}, \quad  h^i_{j+1,k+1} \quad \}$
\EndFor
\EndFor
\end{algorithmic}
after which the pattern repeats for all $N$.

 \begin{figure*}
         \centering
         \includegraphics[width=\textwidth,trim={5cm 0 3.5cm 0},clip]{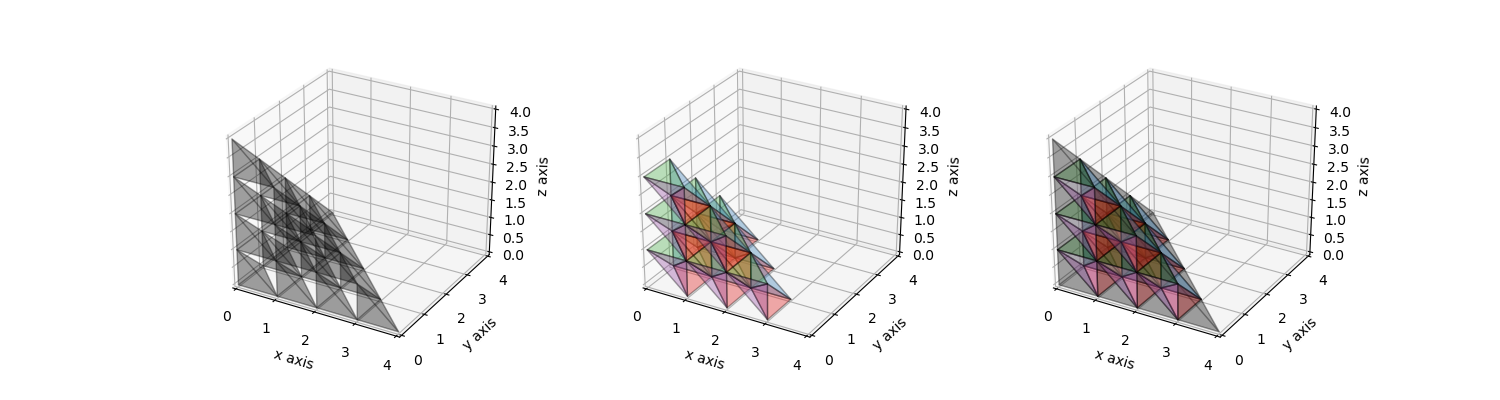}
         \caption{$P_4$ subdivision. }
         \label{fig:p4sub}
\end{figure*}

\clearpage

The author would like to thank L{\'a}szl{\'o} N{\'e}meth for helpful discussions of this problem.


\bibliography{bib}
\bibliographystyle{plain}

\end{document}